**Participatory Design of EHR Components: Crafting Novel Relational Spaces for IT Specialists and Hospital Staff to Cooperate**

Louise Robert[1], Laurine Moniez[1], Quentin Luzurier[2], David Morquin[1]

[1] ERIOS, Espace de Recherche et d'Intégration des Outils Numériques en Santé, CHU de Montpellier
[2] ERIOS, Espace de Recherche et d'Intégration des Outils Numériques en Santé, DEDALUS France

Introduced in the early 2010s, Electronic Health Records (EHRs) are now ubiquitous in hospitals. Despite some undeniable benefits, they remain unpopular among health professionals and present numerous challenges. Studies have documented how EHRs lead to clerical burden, are insufficiently adapted to the reality of healthcare teams' work, and too often result in a loss of meaning and even burnout (Morquin 2019; Gardner et al. 2019; Tai-Seale et al. 2023). This phenomenon is a contributing factor to the ongoing human resources challenges in hospitals.

Our research, situated at the intersection of Health Information Systems studies, Computer Supported Collaborative Work (CSCW), Service Design, and Participatory Design (PD), investigates the transformative impact of involving users in developing new EHRs components within a dedicated space in the hospital. It involves collaboration between a large University Hospital, a University, and a leading European EHR vendor. The research team's interdisciplinary composition ensures a comprehensive perspective in addressing these challenges, comprising a clinician with a PhD in Information Systems Management, four IT Specialists, two Service Designers, and a PhD holder in Linguistics. We have developed two dashboards: Firstly, one that enables all team members involved in the care of patients under the Mental Health Act in psychiatry to coordinate their work (Iso4Psy). Secondly, one centered on individual patient antimicrobial stewardship decision support (AtbViz).

**Crafting Novel Relational Spaces Through The Medium Of Design**

We are examining how integrated design practice can establish essential interaction spaces to overcome the historical challenges associated with the EHRs. Our research focuses on pinpointing the material and interactional elements that enable genuine participatory co-design (hereafter PCD, as used by Melles 2019), positively shifting power dynamics between IT specialists and users, and fostering a more balanced agency in their relationship.

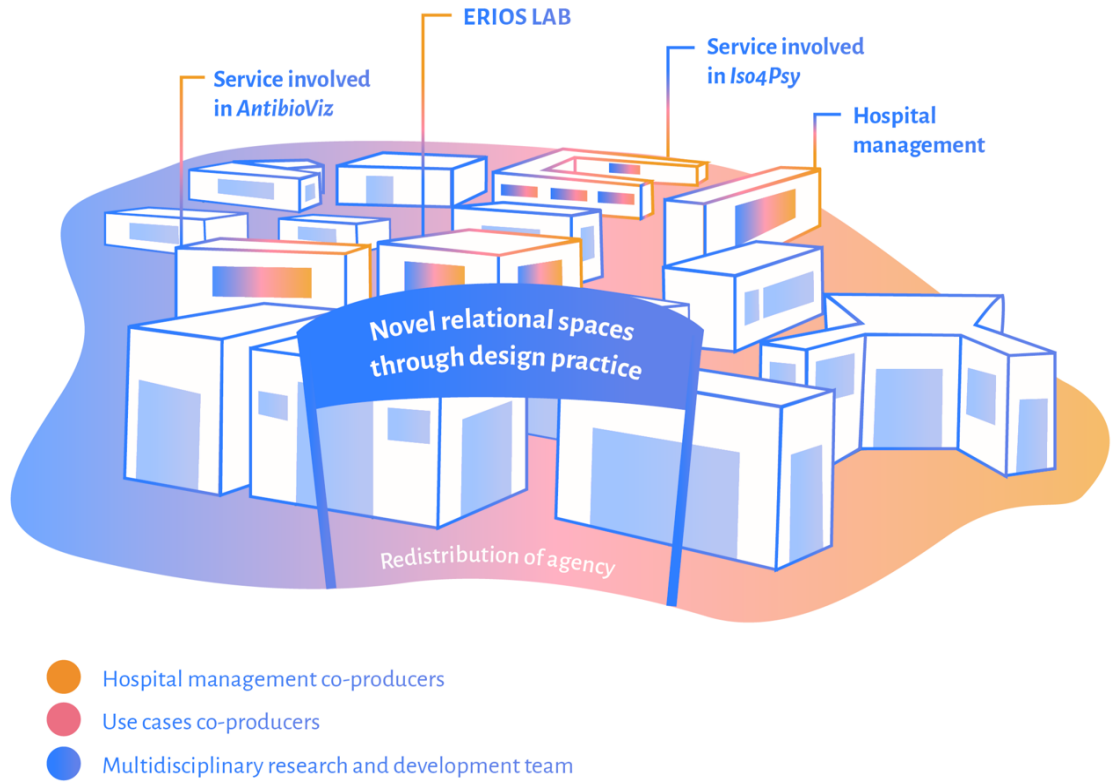

Fig 1. Novel relational spaces through design practice

We employ design tools to create relational areas between stakeholders, following the 'Telling, Making, Enacting' (TME) framework (Brandt, Binder, and Sanders 2013). This approach ensures that our participative design methodology is deeply rooted in practice. In the telling phase, our methodology includes over 100 hours of ethnographic observations and semi-structured interviews per case study as well as narrative workshops.

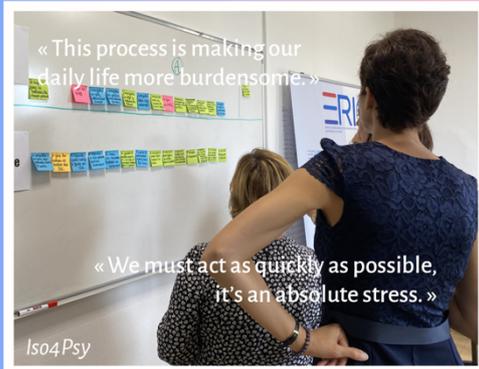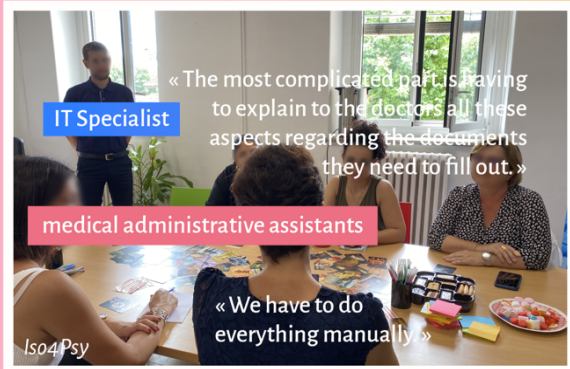

Fig 2. The telling phase

The making phase includes co-design workshops for dashboards development, leading to the production of approximately 100 paper-based prototypes for the two use cases.

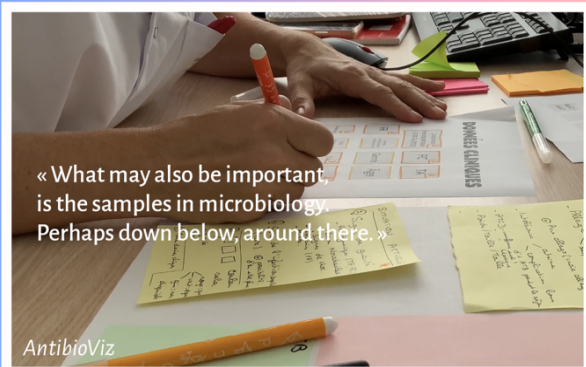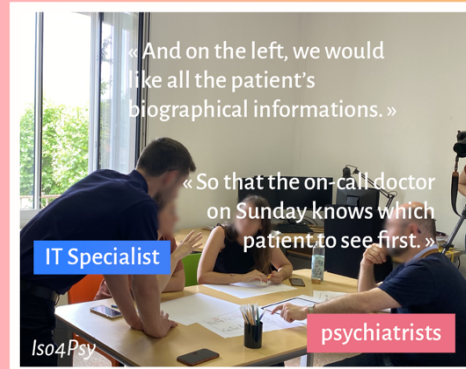

Fig 3. The making phase

In the enacting phase, our evaluation protocol is collaboratively developed and includes the joint creation of scenarios for laboratory tests. This process entails considering the physical space, furniture, and props to authentically recreate situated conditions, encompassing temporal elements like task interruptions or time pressure. In this regard, the practice of service design draws a parallel with theater production (Penin and Tokinwise 2009).

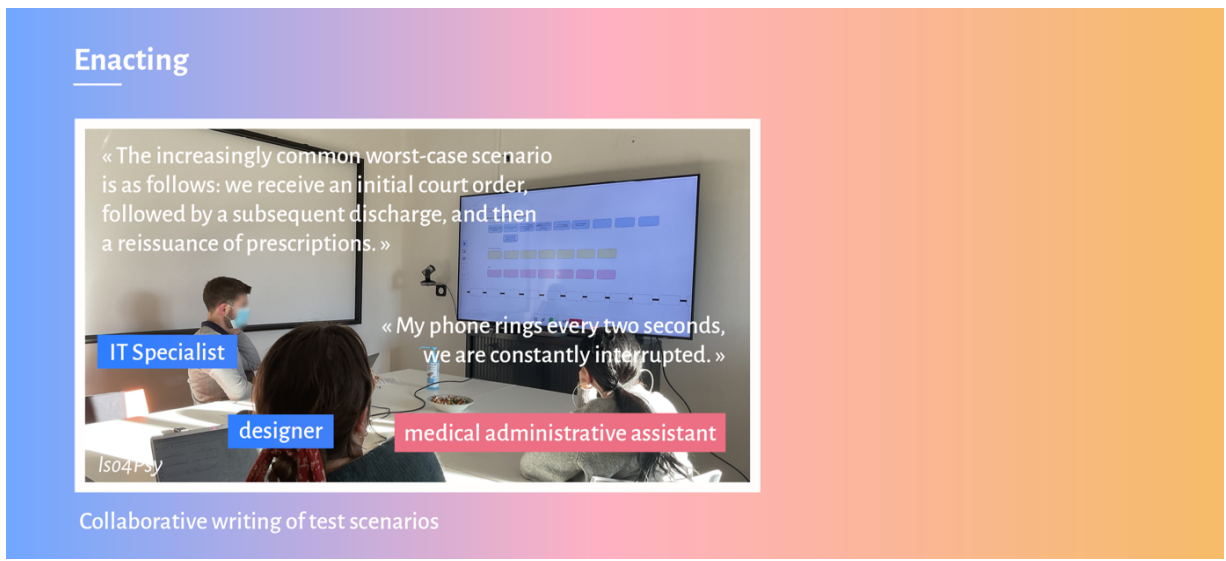

Fig 4. The enacting phase

**Bridging Gaps In EHR Dialogue**

We contend with a lack of clarity surrounding participatory constructs within healthcare (Masterson et al. 2022), a situation that, we find, further deteriorates when attention shifts from patients to Information Systems users, who are rarely involved (Busse et al. 2023). Traditionally, there has been a discrepancy between the availability of medical staff and management for consultation, significantly impacting EHR design. Balka et al. (2018) observed that healthcare managers and system designers often overlook the intricacies and variations of clinical practice. Overturning this perspective is crucial in our project, extending beyond involving doctors to include a diverse array of users.

This approach, however, does not diminish the significance of the perspectives and ideas from the development teams. An essential aspect of service design process is precisely considering the distinctive experiences of both the recipients and the providers (Penin and Tokinwise 2009). When the focus of developing health information systems (HIS) shifts from being strictly confined to the R&D departments of EHR vendors to hospitals, the dynamics undergo a significant transformation. The hospital, as characterized by Balka et al. (2018), is an 'ambiguous territory' marked by complex power dynamics. For development teams it serves as a non-traditional work environment. Our project also strengthens their role by actively engaging them in conversations with users and co-facilitating workshops. Here, too, design practice rebalances power dynamics and agency within these relationships.

**Results & Conclusion**

Our research has successfully transitioned from initial paper prototypes to functional components. Thus, the practice of design, as it becomes progressively integrated into the hospital, induces significant changes. In line with Manzini's argument (2011), the term 'designing for service,' rather than 'designing services,' recognizes that what is being designed is not an end result, but rather a platform for action, with which diverse actors will engage over time. This perspective aligns with the necessity for systemic transformation within the development of HIS where research teams act as facilitators of multidisciplinary innovation via design practice (Minder and Heidemann Lassen 2018). Such a method is a crucial condition for designing an EHR that accurately meets the specific needs of medical professionals and their allies.

In the end, we aim to demonstrate our ability to genuinely renew the approaches used in the development of these software solutions that impact the entire healthcare system. However, the difficulties in establishing these collaborative spaces are numerous and encountered daily. Many competing practical pressures are at play. Creating effective interaction platforms between publishers, the various users, and researchers is a complex and challenging endeavor. It requires not only the alignment of different goals and timelines but also the nurturing of mutual understanding and trust (Metz et al. 2022).

**References**

Balka, Ellen, David Peddie, Serena S. Small, Christine Ackerley, Johanna Trimble, and Corinne M. Hohl. 2018. "Barriers to Scaling up Participatory Design Interventions in Health IT: A Case Study". *Proceedings of the 15th Participatory Design Conference: Short Papers, Situated Actions, Workshops and Tutorial* - Volume 2, 1–5. PDC '18. New York,USA:Association for Computing Machinery.https://doi.org/10.1145/3210604.3210612.

Brandt, Eva, Binder, Thomas and Sanders, Elizabeth. 2013. "Tools and Techniques: Ways to Engage Telling, Making and Enacting". *Routledge International Handbook of Participatory Design.*, 1st ed.

Busse, Theresa Sophie, Chantal Jux, Johannes Laser, Peter Rasche, Horst Christian Vollmar, Jan P Ehlers, and Sven Kernebeck. 2023. "Involving Health Care Professionals in the Development of Electronic Health Records: Scoping Review". *JMIR Human Factors 10 : e45598.* https://doi.org/10.2196/45598.

Gardner, Rebekah L., Emily Cooper, Jacqueline Haskell, Daniel A. Harris, Sara Poplau, Philip J. Kroth, and Mark Linzer. 2019. "Physician Stress and Burnout: The Impact of Health Information Technology". *Journal of the American Medical Informatics Association: JAMIA* 26 (2): 106–14. https://doi.org/10.1093/jamia/ocy145.

Manzini. 2011. "Introduction". *Design for Services*. New York, Routledge.

Minder, Bettina, and Astrid Heidemann Lassen. 2018. "The Designer as Facilitator of Multidisciplinary Innovation Projects". *The Design Journal* 21 (6): 789–811. https://doi.org/10.1080/14606925.2018.1527513.

Morquin, David. 2019. " Comment Améliorer l'usage Du Dossier Patient Informatisé Dans Un Hôpital ? : Vers Une Formalisation Habilitante Du Travail Intégrant l'usage Du Système d'information Dans Une Bureaucratie Professionnelle". *Thèse de Doctorat, Sciences de Gestion*. Université de Montpellier

Penin, Lara, and Tokinwise Cameron. 2009. "The Politics and Theatre of Service Design". *Proceedings of the Conference of the International Association of Societies of Design Research*. IASDR Press, 4327-4338.

Tai-Seale, Ming, Sally Baxter, Marlene Millen, Michael Cheung, Sidney Zisook, Julie Çelebi, Gregory Polston, et al. 2023. "Association of Physician Burnout with Perceived EHR Work Stress and Potentially Actionable Factors". *Journal of the American Medical Informatics Association*, July, ocad136. https://doi.org/10.1093/jamia/ocad136.